\documentclass[conference,letterpaper]{IEEEtran}
\IEEEoverridecommandlockouts

\ifCLASSINFOpdf
\else
\fi
\usepackage[utf8]{inputenc}

\usepackage{units}
\usepackage[tight,footnotesize]{subfigure}
\usepackage{graphicx}
\usepackage{amsmath}
\usepackage{amssymb}

\usepackage{tabularx}
\usepackage{comment}    
\usepackage{setspace}
\usepackage{flushend}
\usepackage{cleveref}
\usepackage{xspace}

\usepackage{acronym}
\acrodef{LEO}{Low Earth Orbit}
\acrodef{LLM}{Large Language Model}
\acrodef{RL}{Reinforcement Learning}
\acrodef{ISL}{inter-satellite link}
\acrodef{DDQN}{Double Deep Q-Network}
\acrodef{LARGE}{LLM-Automated Reward Generation and Enhancement}

\usepackage{xcolor}

\newcommand{\llm}{\ac{LLM}\xspace}
\newcommand{\llms}{\acp{LLM}\xspace}
\newcommand{\rl}{\ac{RL}\xspace}
\newcommand{\leo}{\ac{LEO}\xspace}
\newcommand{\isl}{\ac{ISL}\xspace}
\newcommand{\largereward}{\ac{LARGE}\xspace}
\newcommand{\ddqn}{\ac{DDQN}\xspace}

\renewcommand{\figurename}{Figure}

\newcolumntype{C}[1]{>{\centering\arraybackslash}m{#1}}
\newcolumntype{L}[1]{>{\arraybackslash}m{#1}}

\begin{document}
    \bstctlcite{IEEEexample:BSTcontrol}
	%





	%



\title{LLM-Driven Automated Reward Design for Reinforcement Learning-Based Routing in \\LEO Satellite Networks}%
    \author{\IEEEauthorblockN{Walter~P.~Casas, Nelson~L.~S.~da~Fonseca, and Carlos~A.~Astudillo}
    \IEEEauthorblockA{Institute of Computing - State University of Campinas, Brazil \\ E-mail: w187990@dac.unicamp.br, nfonseca@ic.unicamp.br, and castudillo@unicamp.br}}
    
	\maketitle

\begin{abstract}

Routing in \leo satellite networks is challenging due to highly dynamic topologies and spatio-temporal network conditions. \rl has emerged as a promising approach for adaptive routing; however, its performance critically depends on reward function design, which must balance objectives such as goodput and end-to-end delay. In practice, reward design remains a complex manual process requiring significant domain expertise and extensive trial-and-error. Recent works have explored \llms for automated reward design, but their application to highly dynamic systems such as \leo satellite networks remains largely unexplored. We propose \acs{LARGE}, a framework that automates reward design for \rl-based routing by combining \llm-driven generation with iterative simulator-in-the-loop evaluation. LARGE generates an initial reward from \llm prior knowledge and iteratively refines it using simulation feedback. This loop enables exploration of diverse reward formulations while aligning them with network objectives. Results show that LARGE improves reward quality within a few iterations through feedback-driven refinement. Across different backbones, the framework achieves performance comparable to an expert-designed baseline, with the best-performing configuration reaching goodput within approximately 3\% of the baseline and slightly lower end-to-end delay, without manual reward engineering. These results indicate that effectiveness emerges from the iterative feedback-driven process enabled by LARGE, highlighting the potential of framework-driven \llm-in-the-loop optimization for \rl-based routing in dynamic satellite networks.

\end{abstract}
	


%

    \acresetall

	\section{Introduction}
\label{sec:Introduction}

\leo satellite constellations are becoming a key component of future global connectivity, particularly in the context of 6G networks, which target ubiquitous, low-latency, and high-capacity communication services. Operating at altitudes typically ranging from 400 to 1200~km, \leo systems offer significantly lower propagation latency compared to geostationary satellites~\cite{koosha2025comprehensive}. However, their movement at speeds of approximately 7.6 km/s causes the network topology to change constantly. In such a dynamic environment, routing requires selecting sequences of satellite nodes and inter-satellite links to forward data units, from source to destination. The resulting time-varying \isl availability, frequent link disruptions, and changing resource conditions make traditional routing algorithms less effective~\cite{shi2024leo,sciencedirect2025routing}.

\rl has recently emerged as a promising solution for these challenges, allowing satellites to make distributed routing decisions with only partial knowledge of the environment. However, the effectiveness of any \rl agent critically depends on the design of the reward function, which guides the policy toward desired behaviors. A poorly specified reward can lead to slow convergence, inefficient, or unintended strategies. Moreover, designing an appropriate reward requires expert knowledge and repeated trial-and-error experimentation~\cite{kwon2023reward,booth2023perils}. This motivates the need for automated reward design methods that address the specific challenges of \leo satellite routing.

Recent work has demonstrated that LLMs can automate reward function design across multiple domains. Early approaches leveraged LLMs to evaluate agent behavior and provide reward signals during RL training~\cite{kwon2023reward} or to generate reward functions from language for direct optimization in robotic control~\cite{yu2023language}. Subsequent efforts moved toward generating executable reward code, with Text2Reward~\cite{xie2024text2reward} producing reward functions from task descriptions using structured environment representations and optional human feedback, and~\cite{jin2024reward} introducing intermediate reward components to improve logical structure in robotic assembly tasks. However, these methods rely on human input or annotated datasets, which may limit scalability and generalization. 

More recent frameworks close the loop between LLMs and RL agents by iteratively refining rewards using training feedback: CARD~\cite{sun2025large} updates reward code based on trajectory-derived success–failure preferences, AutoReward~\cite{han2024autoreward} leverages simulation metrics such as success rates in autonomous driving, and~\cite{baek2024chatpcg,zhu2025llm} employs Chain-of-Thought reasoning to iteratively decompose and refine reward design. Despite these advances, such approaches primarily target robotic or game environments, where feedback signals are well defined and closely aligned with task objectives. 

In the context of networking, recent surveys~\cite{zhou2024large,jiang2026comprehensive} highlight the growing role of LLMs in network management and RL optimization. Building on this trend, Zheng et al.~\cite{zheng2026large} propose a taxonomy of LLM roles in 6G RL optimization and explore UAV–satellite scenarios, while~\cite{cai2026large} uses LLMs to design enriched reward signals and process environment feedback for UAV data collection. Nevertheless, these works neither consider iterative reward refinement across training runs nor address the highly dynamic, topology-driven characteristics of LEO satellite networks. To the best of our knowledge, no prior work proposes a closed-loop framework in which an LLM autonomously generates and refines reward functions based on structured network metrics for DDQN-based routing in LEO satellite constellations, which is the main focus of this paper.


Thus, this work proposes \largereward, a \llm-based framework that generates an initial reward function from a problem description and iteratively refines it using feedback from network state and routing performance. Rather than relying on manual reward engineering, \largereward explores whether structured performance feedback can guide an LLM toward reward functions that are competitive with expert-designed alternatives in dynamic \leo satellite routing. The main contributions of this work are as follows. First, we introduce \largereward, an automated framework for reward generation and refinement in \rl-based routing for \leo satellite networks. Second, we design a feedback-driven optimization loop in which the reward function is iteratively revised based on observed network conditions and policy performance. Third, we evaluate the proposed approach through simulation in a dynamic routing environment and show that LLM-generated rewards can achieve performance comparable to an expert-designed baseline within a few refinement iterations, with the best-performing configuration reaching goodput within approximately 3\% of the baseline and slightly lower delay, without requiring manual reward engineering.

    \section{Problem Formulation}

Consider an agent operating in a Markov decision process (MDP) defined by $M = (\mathcal{S}, \mathcal{A}, \mathcal{P})$, where $\mathcal{S}$ is the state space, $\mathcal{A}$ is the action space, and $\mathcal{P}$ is the transition function. The space of candidate reward functions is denoted by $\mathcal{R}$, where each $r \in \mathcal{R}$ maps states and actions to scalar values guiding the learning process. A \rl algorithm operating on $M$ is denoted by $\mathcal{T}_M(r)$, which takes a reward function $r$ as input and produces a policy $\pi \in \Pi$. The fitness function $F: \Pi \to \mathbb{R}$ evaluates the quality of a learned policy $\pi$.

Based on these definitions, the \emph{reward design problem}~\cite{singh2009rewards} consists of finding a reward function $r^* \in \mathcal{R}$ such that the induced policy maximizes the fitness, as given in Eq.~\ref{eq:rdp}.

\begin{equation}
    r^* = \arg\max_{r \in \mathcal{R}} 
    F\!\left(\mathcal{T}_M(r)\right)
    \label{eq:rdp}
\end{equation}

The \emph{reward generation problem} consists of finding a reward function $\hat{r}$ that approximates the optimal reward $r^*$. In our context, we employ a pretrained \llm $\mathcal{G}_\theta$, where $\theta$ denotes the model parameters, together with a prompt $p$ that describes the network task and optimization objectives. The model generates a candidate reward function as $\hat{r} = \mathcal{G}_\theta(p)$, which is then used by the training algorithm $\mathcal{T}_M(\hat{r})$ to produce a policy $\hat{\pi}$. The fitness function $F$ evaluates $\hat{\pi}$ using network performance metrics obtained via simulation.

    \section{LARGE: LLM-Automated Reward Generation and Enhancement}
\label{sec:proposal}

This section introduces \largereward, a framework based on \llms for the automated generation and optimization of reward functions for RL-based routing in LEO satellite network constellations. LARGE operates in two stages: \textit{Cold Start Reward Generation}, which produces an initial reward function, and \textit{Iterative Reward Improvement}, which iteratively refines the reward function until it converges toward a predefined objective, expressed as a target goodput value. The framework relies on three \llm agents. The \textbf{Metrics Interpreter Agent} analyzes network performance metrics and translates them into a structured prompt. The \textbf{Reward Design Agent} generates reward function definitions based on its internal knowledge and the provided context. The \textbf{Code Generator Agent} implements those definitions as executable code, verifying that all required variables are available in the Network Simulation Environment.

\begin{figure}[!t]
    \centering
    \includegraphics[width=\linewidth]{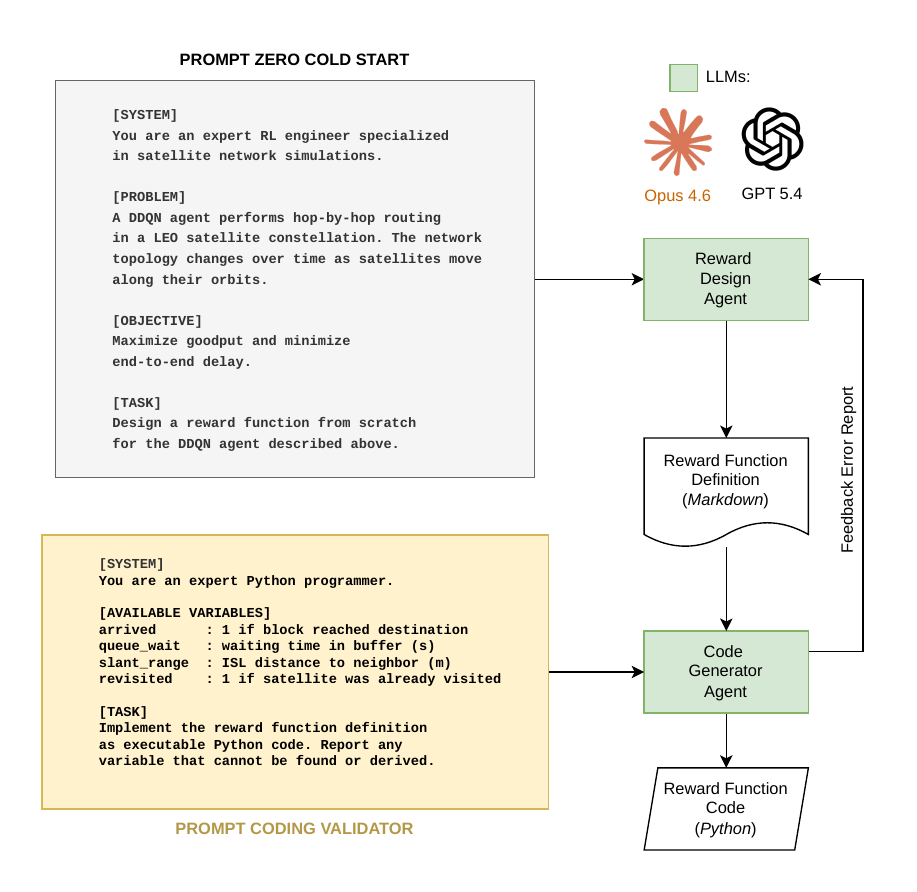}
    \caption{Cold Start Generation stage of LARGE. The Reward Design agent generates a reward function from a problem description prompt; the Code Generator implements it with available variables and, if unsuccessful, requests a revision.}
    \label{fig:zerocold}
\end{figure}

\begin{figure*}[!t]
    \centering
    \includegraphics[width=\linewidth]{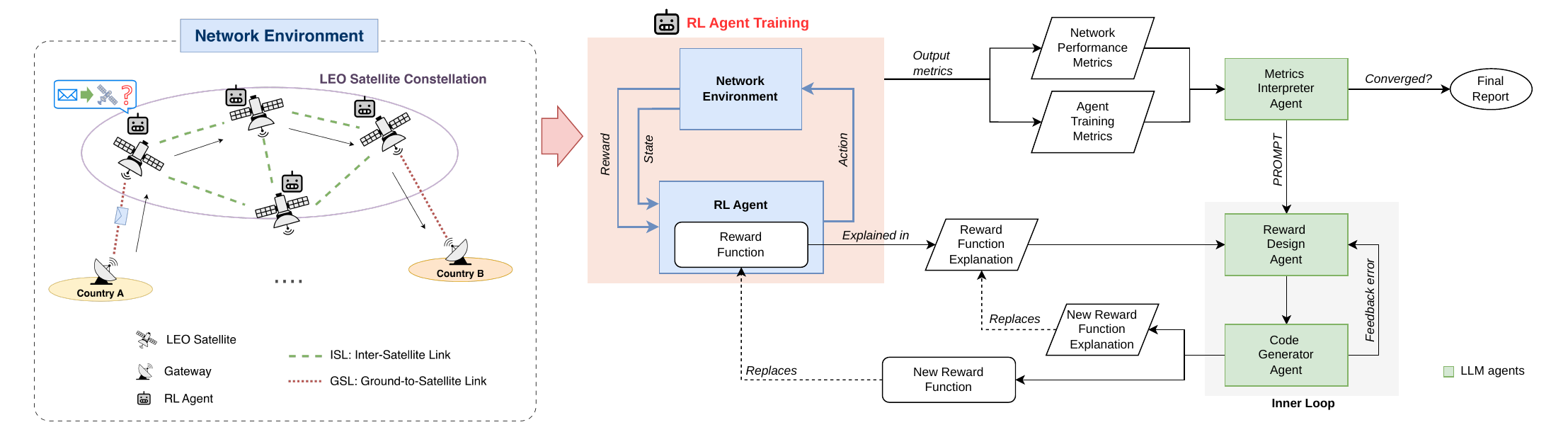}
    \caption{Iterative Reward Improvement stage of LARGE. The Metrics Interpreter Agent analyzes the performance metrics returned by the Network Simulation Environment, the Reward Design Agent generates a new reward function definition, and the Code Generator Agent implements it. This loop continues until the convergence criterion is met.}
    \label{fig:framework}
\end{figure*}

\subsection{Cold Start Reward Generation}
The goal of this stage is to produce an initial reward function without bias toward environment specific variables. The Reward Design Agent receives a prompt describing the routing problem and the optimization objective and generates a reward function definition, which is passed to the Code Generator Agent. The Code Generator Agent verifies whether each variable is available in the Network Simulation Environment, either directly or through derivation from other variables. If a variable is not available for implementation in Python, the Code Generator Agent returns the definition along with a report listing the missing variables and requests a revised version. This loop continues until a valid definition is obtained, preventing the Reward Design Agent from relying only on a limited set of variables and encouraging the Code Generator Agent to derive the required ones from available knowledge. Once all variables are confirmed, the Code Generator Agent implements the reward function in code and deploys it in the Network Simulation Environment, as shown in Fig.~\ref{fig:zerocold}. At this point, the process proceeds to the second stage of \largereward.

\subsection{Iterative Reward Improvement}
In this stage, the process, illustrated in Fig.~\ref{fig:framework}, begins with training the \rl agent in the Network Simulation Environment using the reward function implemented in the previous stage. Once the training phase is completed, the environment returns a set of network performance metrics, including goodput in Mbps, path stretch, and end to end delay in ms, along with training indicators such as cumulative reward and training loss. The Metrics Interpreter Agent evaluates whether the convergence criterion, defined according to state of the art metrics, is satisfied; if so, the process stops and the trained agent is ready for evaluation. Otherwise, it converts the metrics obtained at the end of training into a structured prompt that highlights areas for improvement, reinforces the optimization objective, and indicates whether the reward function improved or degraded compared to the previous run, and sends it to the Reward Design Agent for refinement.

The structured prompt is then passed to the Reward Design Agent, initiating an inner loop between the Reward Design Agent and the Code Generator Agent, similar to that of the previous stage. The Reward Design Agent interprets the feedback from the previous training run and proposes improvements by defining a new reward function, along with a brief rationale. This definition is sent to the Code Generator Agent, which verifies that all required variables are available in the environment and attempts to implement the reward function in Python. If implementation fails due to missing variables, the Code Generator Agent returns an error specifying the unavailable elements and requests a revised proposal. This interaction continues until a valid implementation is obtained using only available variables. Once successful, the Code Generator Agent generates a markdown document explaining the reward function and the reasoning behind its design. The newly implemented reward function replaces the previous one, the \rl agent is retrained under the same conditions, and the cycle repeats until the convergence criterion is satisfied.

This iterative process produces the final optimized reward function. The experimental setup and baseline used for its evaluation are described in the following section.

	\section{Performance Evaluation}
\label{sec:perfromance-evaluation}

\subsection{Network Simulation Environment}

The experiments are conducted using the open-source routing simulator proposed in \cite{lozanocuadra2024simulator}, which models packet routing in \leo satellite constellations. The simulator represents a realistic scenario where ground gateways aggregate terrestrial traffic from nearby users into blocks of $B=64{,}800$ bits with the same destination and inject them into the constellation as packets, which are then routed through a dynamic network of satellites until they reach their destination gateway. The environment is implemented using an event-based discrete-time approach and built on a dynamic time-varying graph, where nodes represent satellites and gateways, and edges correspond to communication links, including inter-satellite links (ISLs) and ground-to-satellite links (GSLs). This allows the simulator to model traffic generation, packet forwarding, queuing, transmission, and propagation dynamics.

In this setting, the routing problem consists of deciding, at each satellite, the next hop to which a packet should be forwarded among the currently available neighboring nodes. Thus, satellites act as routers within the \leo constellation, and the selected sequence of next-hop decisions determines the end-to-end path followed by each packet. The \rl framework follows a multi-agent setting, where each satellite acts as an independent agent responsible for making local routing decisions based on partial network information. The learning process includes two phases: during the online phase, agents learn from interactions with the environment through exploration, while during the offline phase, pre-trained models are deployed for decision making. The agents are trained using a \ddqn algorithm, which is used in all experiments. Since the focus of this work is on reward optimization, we primarily analyze the behavior of the agents during the training phase.

\subsection{Experimental Setup}
The proposed framework is evaluated on the Kepler constellation, which consists of 140 satellites distributed across seven orbital planes at an altitude of 600~km. We use a fixed constellation to isolate the effect of reward optimization from changes in orbital topology, link dynamics, and path-length distributions. This controlled setting enables a direct comparison between reward functions while keeping the routing environment unchanged. The hyperparameters of the \ddqn agent are kept fixed across all experiments and set to the default values provided by the simulator~\cite{lozanocuadra2024simulator}, ensuring that only the reward function varies. As a baseline, we use the reward function provided by the original simulator, designed by domain experts. The framework is evaluated using two generative \llm backbones: GPT-5.4, developed by OpenAI~\cite{openai_gpt54}, referred to as LARGE-GPT, and Claude Opus~4.6, developed by Anthropic~\cite{anthropic_opus46}, referred to as LARGE-Opus. Each backbone is used consistently across all \llm agents in the pipeline and evaluated independently until the convergence criterion is satisfied.

The evaluation follows a three-phase protocol. First, during the \textit{search phase}, the proposed LARGE framework is used to optimize the reward function. Each candidate reward function generated by LARGE is evaluated by training the \ddqn agent for 0.2 seconds, producing approximately 70,000 hop-level reward events and 35,000 training steps. This short evaluation window provides sufficient signal to estimate early convergence behavior while keeping the \llm-in-the-loop reward search computationally feasible. The \largereward search loop terminates once a candidate reward function achieves a goodput higher than that of the baseline reward, and the selected reward function is then carried forward to the subsequent validation phases. Then, during the \textit{training phase}, a new agent is trained from scratch for a full second using only the selected reward function, with performance recorded at checkpoints every 0.2 seconds. This phase verifies that the selected reward captures genuine routing objectives rather than exploiting the short search horizon used during \largereward optimization. Finally, during the \textit{inference phase}, the trained policy is deployed without further learning for 12 seconds, during which satellite positions are updated over time. This phase serves as the primary benchmark for assessing generalization under realistic dynamic conditions. To account for the stochastic nature of the training process, all results are reported as mean $\pm$ standard deviation over 10 independent runs with different random seeds.

\subsection{Evaluation Metrics}
The performance of the \rl agent is evaluated using the following network metrics:

\begin{itemize}
\item \textbf{Path stretch}: ratio between the hop count of the path followed by a routed packet and the hop count of the shortest path computed by Dijkstra's algorithm. A value of 1 indicates an optimal route in terms of hop count.
\item \textbf{Goodput}: amount of data successfully delivered to its destination per unit of time, measured in Mbps.
\item \textbf{Delay}: average time elapsed from the moment a traffic block of $B=64{,}800$ bits is generated at the source gateway until it reaches its destination gateway, measured in milliseconds.
\end{itemize}

\subsection{Results}

\begin{figure}[!b]
    \centering
    \includegraphics[width=0.99\linewidth]{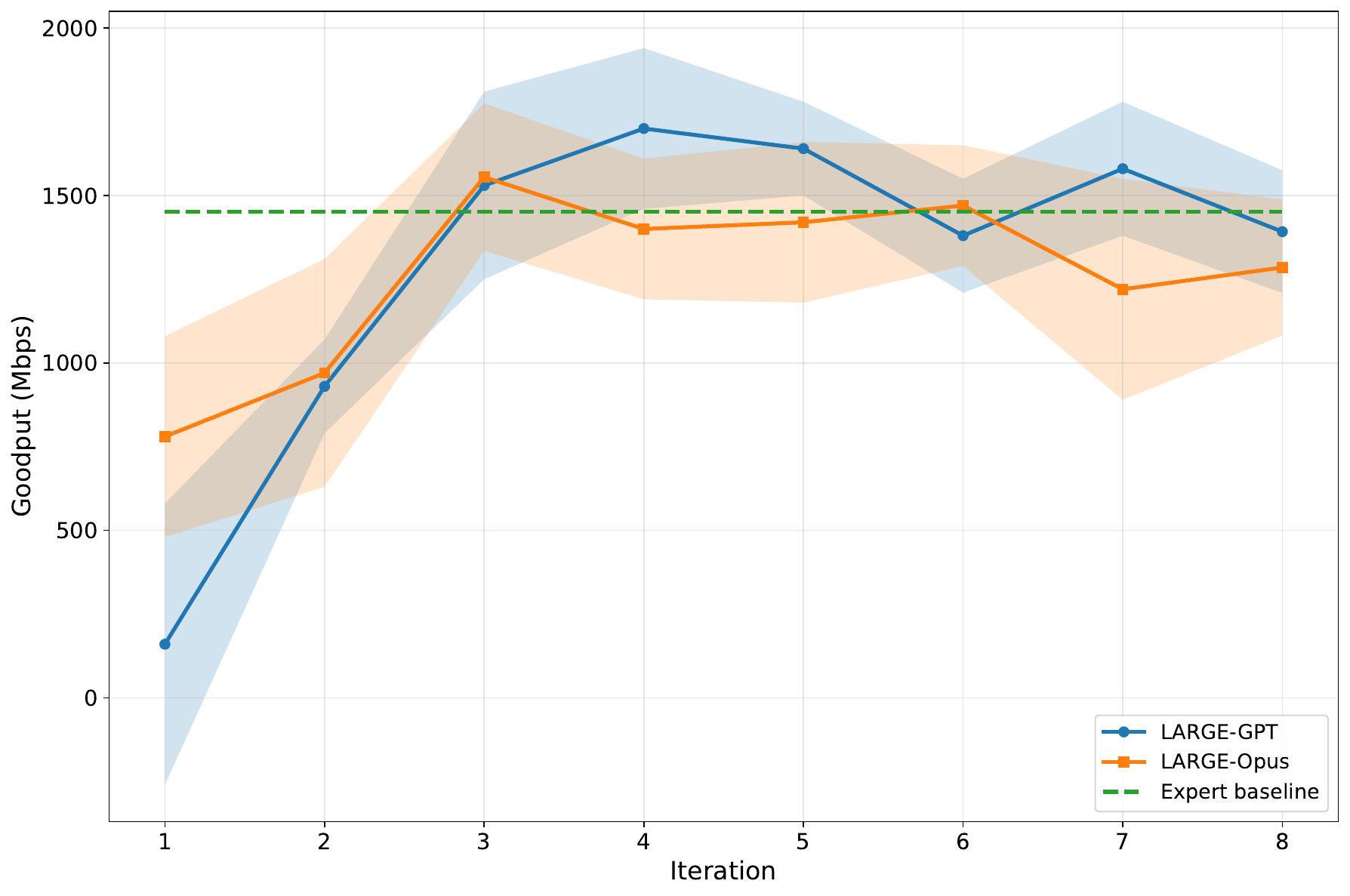}
    \caption{Goodput achieved by LARGE across iterations. The dashed line represents the reward baseline designed by experts. LARGE converges once the generated reward function surpasses the baseline performance.}
    \label{fig:goodput}
\end{figure}

When \largereward is applied during the search phase, the framework reaches the goodput-based stopping criterion by the third iteration, using only a training window of 0.2 seconds per candidate reward, as shown in Fig.~\ref{fig:goodput}. The same figure also illustrates the cold-start behavior of the reward generation process: the reward proposed in the first iteration is not sufficient to satisfy the criterion, which motivates the need for iterative refinement. Once this refinement starts, the feedback from the previous training run leads to improved reward proposals, which is reflected in the increase in goodput. In addition, after the stopping criterion is reached, subsequent iterations do not produce substantial performance gains. This behavior is associated with more conservative reward proposals, where the modifications mainly consist of small changes to the coefficient values.

\begin{figure}[!t]
    \centering
    \includegraphics[width=\linewidth]{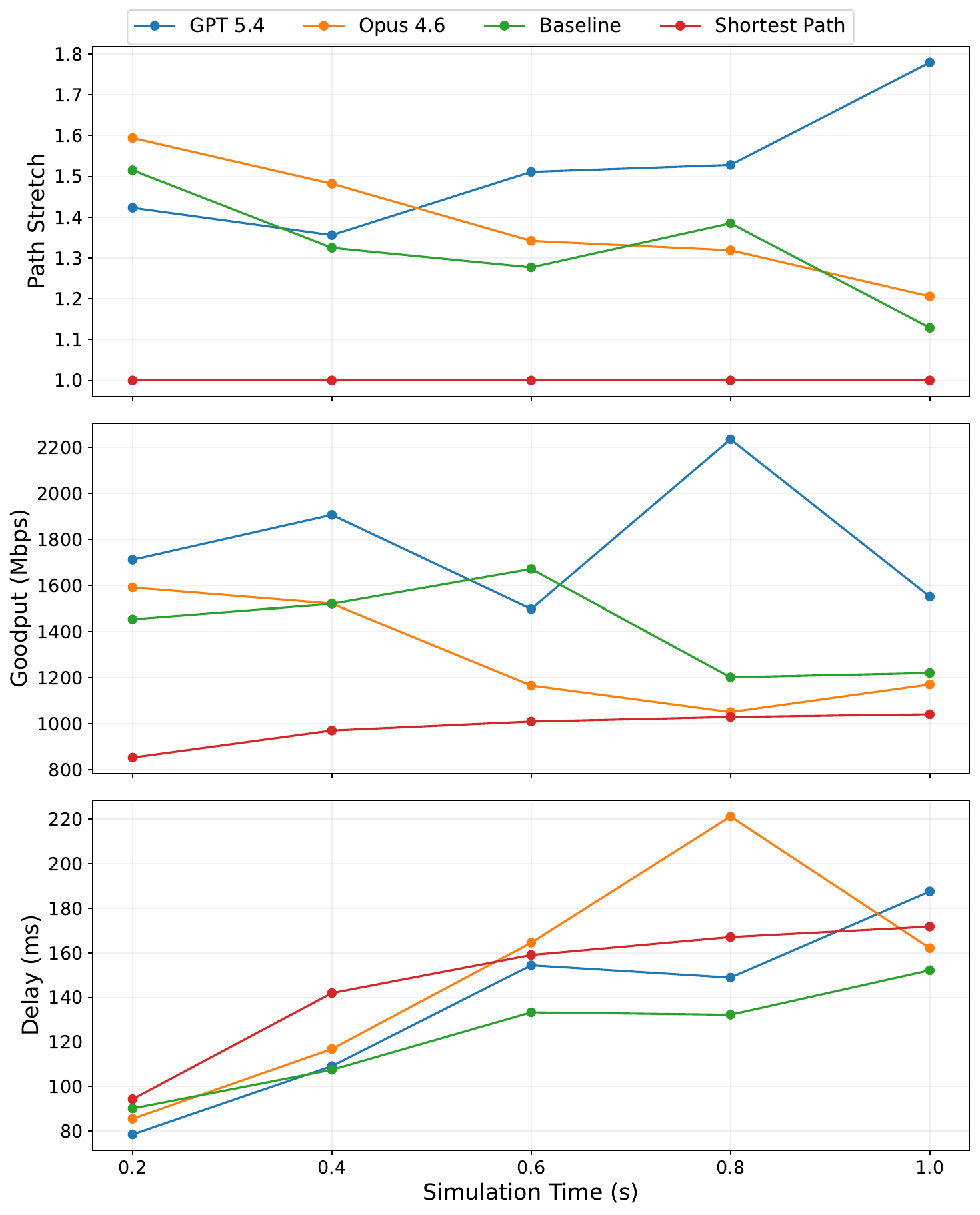}
    \caption{Network performance metrics across LARGE iterations and training 
    checkpoints for GPT-5.4 and Claude Opus 4.6.}
    \label{fig:result}
\end{figure}

Based on this behavior, the best-performing reward functions obtained during the search phase are selected for further analysis. These rewards are then used in the second phase, where new \ddqn agents are trained from scratch for a full second, with performance recorded at checkpoints every 0.2 seconds, as shown in Fig.~\ref{fig:result}. This stage evaluates whether the rewards that satisfied the stopping criterion in the short 0.2-second search window remain effective when training is extended. The results show that both LARGE-GPT and LARGE-Opus maintain competitive performance throughout the full training interval. LARGE-GPT achieves the highest goodput in most checkpoints, although this comes with a higher path stretch, especially toward the end of training. In contrast, LARGE-Opus exhibits a more conservative behavior, with lower path stretch and performance closer to the expert baseline across the evaluated metrics. Overall, both rewards preserve the gains observed during the search phase and consistently achieve higher goodput than the shortest-path routing strategy, confirming that the selected rewards capture useful routing behavior beyond the initial search window.

Finally, the agents trained during the second phase are evaluated in a 12-second inference scenario, where no further learning is performed and satellite positions are updated over time. This stage assesses whether the rewards selected during the short search phase and validated during the one-second training phase produce consistent routing behavior under a longer dynamic evaluation. The results, reported in Table~\ref{tab:results}, show that the generated rewards remain competitive with the expert-designed baseline. LARGE-Opus achieves the closest overall performance to the baseline, with comparable goodput, slightly lower delay, and a similar path stretch. LARGE-GPT obtains lower goodput and a higher path stretch, indicating a less stable transfer from training to inference, although its delay remains close to the baseline. Overall, these results suggest that the proposed framework can generate rewards that preserve meaningful routing behavior beyond the short optimization window, with LARGE-Opus showing the most consistent generalization across metrics.

\begin{table}[]
\centering
\caption{Performance comparison between LLM-generated rewards and the expert 
baseline during the inference phase.}
\label{tab:results}
\small
\setlength{\tabcolsep}{4pt} 
\begin{tabular}{lccc}
\hline
\textbf{Method} & 
\textbf{\shortstack{Goodput\\(Mbps)}} & 
\textbf{\shortstack{Delay\\(ms)}} & 
\textbf{\shortstack{Path\\stretch}} \\
\hline
Baseline     & $1451.62 \pm 131.67$  & $85.65 \pm 2.44$ & $1.464 \pm 0.066$  \\
LARGE-GPT    & $1324.03 \pm 237.34$  & $88.41 \pm 2.32$ & $1.569 \pm 0.088$  \\
LARGE-Opus   & $1409.56 \pm 133.10$  & $85.13 \pm 3.03$ & $1.486 \pm 0.034$ \\
\hline
\end{tabular}
\end{table}


\subsection{Reward Analysis}
\label{sec:reward_analysis}

To better understand the type of reward structures produced by \largereward, we analyze the reward function generated by LARGE-Opus and compare it with the expert-designed baseline. The baseline reward follows a relatively compact design, mainly combining a distance-based term, a queue-based term, and fixed penalties or bonuses for special events such as delivery, unavailable links, and loop formation. In contrast, the generated reward adopts a more structured goodput-oriented formulation. Rather than evaluating only the selected hop in isolation, it compares the chosen action against the set of currently available neighboring satellites and assigns reward components based on local ranking criteria.

The generated reward introduces several mechanisms that are not present in the expert baseline. First, it includes local ranking terms that favor neighbors with higher data rate and better progress-time efficiency. Second, it modifies the queue component by relating the observed queueing delay to the best available service time, making the penalty depend on the relative quality of the selected hop rather than only on the absolute queueing time. Third, it adds an explicit hop cost, discouraging unnecessarily long routes. Finally, it strengthens loop avoidance by penalizing repeated visits and ping-pong behavior, using satellite identifiers to detect revisits more robustly.

These changes indicate that \largereward is able to produce reward functions that go beyond simple coefficient tuning and introduce additional routing-aware structure. However, the inference results also show that a more expressive reward does not necessarily lead to uniformly better generalization. In particular, the goodput-oriented design of LARGE-GPT can increase throughput during training, but it also tends to produce larger path stretch and less stable inference performance than LARGE-Opus. Therefore, the reward analysis suggests that the main value of \largereward is not only in improving a single metric, but also in exposing alternative reward-design trade-offs between throughput, delay, and route efficiency. Table~\ref{tab:reward} summarizes the main differences between the expert baseline and the reward generated by LARGE-Opus.

\begin{table}[!b]
\centering
\caption{Main structural differences between the expert baseline reward and the reward generated by LARGE-Opus.}
\label{tab:reward}
\small
\setlength{\tabcolsep}{3pt}
\renewcommand{\arraystretch}{1.15}
\begin{tabularx}{\columnwidth}{p{0.26\columnwidth}XX}
\hline
\textbf{Aspect} & \textbf{Baseline} & \textbf{LARGE-Opus} \\
\hline
Action evaluation & Selected hop only & Local neighbor ranking \\
Rate awareness & Not explicit & Data-rate ranking term \\
Efficiency term & Distance-based & Progress-time efficiency \\
Queue penalty & Absolute queue time & Relative service-time penalty \\
Hop cost & Not explicit & Explicit per-hop cost \\
Loop handling & Fixed revisit penalty & Revisit and ping-pong penalties \\
Delivery reward & Mostly fixed & Scaled with additional bonuses \\
Invalid action & Fixed penalty & Stronger penalty \\
\hline
\end{tabularx}
\end{table}

\subsection{Discussion}

The results show that \largereward can generate reward functions that remain competitive with the expert-designed baseline after a short feedback-driven search process. During the search phase, both LARGE-GPT and LARGE-Opus satisfy the goodput-based stopping criterion within a few iterations, and the subsequent training and inference phases indicate that the selected rewards preserve meaningful routing behavior beyond the 0.2-second search window.

The two backbones exhibit different trade-offs. LARGE-Opus produces a more aggressive and structurally richer reward, but this additional complexity does not translate into uniformly better inference performance, as it shows lower goodput and higher path stretch than the expert baseline in the 12-second evaluation. In contrast, LARGE-GPT produces a more conservative reward that remains closer to the baseline across metrics, achieving comparable goodput, slightly lower delay, and similar path stretch. These results suggest that \largereward is useful not only for automating reward design, but also for exposing alternative trade-offs between goodput, delay, and route efficiency.

	\section{Conclusion}
\label{sec:Conclusion}

In this work, we proposed \largereward, an \llm-based framework for automated reward generation and optimization in \rl-based routing for \leo satellite networks. The framework uses three \llm agents within two nested loops: an outer loop that refines candidate rewards using simulator feedback and network metrics, and an inner loop that validates their executable implementation. Experimental results show that \largereward can transform initially imperfect proposals into competitive rewards within a few refinement iterations, without manual reward engineering. The generated rewards satisfy the goodput-based stopping criterion during search and preserve meaningful routing behavior during longer training and inference, showing that the benefit comes from the closed-loop interaction between reward design, code validation, simulation, and feedback-driven refinement.

By connecting \llm-based reward design with simulator-based evaluation, \largereward provides a practical mechanism for translating high-level network objectives into executable reward functions for dynamic communication environments. The results also show that different \llm backbones lead to routing trade-offs between goodput, delay, and route efficiency. Although this work focuses on a controlled Kepler constellation scenario to isolate reward optimization, future work will extend the evaluation to additional constellation architectures, traffic loads, gateway deployments, and longer inference horizons. Further directions include robust multi-objective stopping criteria, prompt sensitivity analysis, and fine-tuned \llm{}s to improve convergence speed and reward quality.

\section*{Acknowledgments}
This work was partially funded by the INCT of Intelligent Communications Networks and the Internet of Things (ICoNIoT) funded by CNPq (process 405940/2022-0) and CAPES (Finance Code 88887.954253/2024-00), and CNPq grant 403979/2023-4.

	\bibliographystyle{IEEEtran}
	\bibliography{references}
	
\end{document}